# Privacy of Fitness Applications and Consent Management in Blockchain

May Alhajri*¶, Ahmad Salehi S.*§, and Carsten Rudolph*
*Monash University, Melbourne, Australia, §RMIT University, Melbourne, Australia,
¶King Faisal University, Al-Ahsa, Saudi Arabia
{may.alhajri,ahmad.salehishahraki,carsten.rudolph}@monash.edu
{ahmad.salehi.shahraki@rmit.edu.au},{mnalhajri@kfu.edu.sa}

## ABSTRACT

The rapid advances in fitness wearable devices are redefining privacy around interactions. Fitness wearables devices record a considerable amount of sensitive and private details about exercise, blood oxygen level, and heart rate. Privacy concerns have emerged about the interactions between an individual's raw fitness data and data analysis by the providers of fitness apps and wearable devices. This paper describes the importance of adopting and applying legal frameworks within the fitness tracker ecosystem. In this review, we describe the studies on the current privacy policies of fitness app providers, heuristically evaluate the methods for consent management by fitness providers, summarize the gaps identified in our review of these studies, and discuss potential solutions for filling the gaps identified. We have identified four main problems related to preserving the privacy of users of fitness apps: lack of system transparency, lack of privacy policy legibility, concerns regarding one-time consent, and issues of noncompliance regarding consent management. After discussing feasible solutions, we conclude by describing how blockchain is suitable for solving these privacy issues.

## 1 INTRODUCTION

The recent proliferation of wearable fitness technologies (e.g., smartwatches, fitness trackers) enables people to generate quantities of valuable data about their health [25]. These wearable devices capture various data about physical activity and health-related data including information about sleep quality and quantity, number of steps and distance, and blood pressure [42]. In the context of health, the use of these technologies supports individuals, physicians, and clinical researchers by enabling accurate diagnoses and evidence of patients' self-reported data [20, 41]. However, as occurs with the adoption of any technology, some challenges have emerged.

The main issue surrounding the data collected by wearable technologies relate to the ownership and privacy of personal health information. Many users have limited or no control over their data, knowledge about the purposes for which the data are used, or their rights [25]. Privacy can be defined as the ability of users to control the sharing, collecting, and processing of their data with other entities [36, 39]. Privacy is an essential element for protecting an individual's data, and this concept creates a significant issue for service providers [49]. For example, in the context of data sharing, data requestors and processors must comply with the relevant rules, such as the European Union's General Data Protection Regulation (GDPR) [1], which makes obtaining the subject's consent essential to collecting, processing, and sharing their data. The GDPR [1] imposes on data processors (e.g., fitness app providers) data protection requirements that empower users to take control of their fitness data.

Researchers have highlighted privacy concerns around the use of fitness apps [2, 23–25, 30–32, 35, 47], which this review covers in detail. Although these studies have thoroughly analyzed the privacy issues related to the use of fitness apps, none were found to have addressed the privacy issues related to fitness apps in terms of dynamic consent management that uses a transparent, legally compliant system. Considering these challenges, studies in various domains have shown empirically that blockchain technology is suitable for mitigating issues related to the preservation of privacy by managing the individual user's consent using a smart contract [7, 16, 21, 22, 28, 38].

Blockchain and smart contracts have been used for consent management in different contexts and should provide a system that is both human centric and legally compliant. The design of the existing systems enables the management of user consent both before and after collecting, processing, and sharing data between the data subject and data requester. Smart contracts can be leveraged to manage user consent by allowing the user to grant or deny requests for access to individual health data. This is important because the use of fitness data can play a significant role in improving the functions of various entities, including health-care providers and other organizations [40, 44]. For example, these data may be used by clinicians to improve diagnoses, by workplace wellness programs [13], and to reduce health insurance costs [12], all of which may be vulnerable to unlawful disclosure of health data without subject's consent.

As first step in our analysis, we reviewed the relevant data protection legal frameworks and data categorizations used by wearable fitness devices. We used this information to identify concerns about privacy issues around the use of fitness apps reported in the literature that has assessed fitness apps and privacy issues. We aimed to identify the privacy issues related to the consent-management practices of fitness apps by identifying the requirements in the existing legal frameworks for protecting fitness data and to compare the current privacy policies of fitness app providers with these level requirements.

### 1.1 Our contribution

The main research contribution of our paper are as follows.

a) to identify and understand the importance of adopting legal frameworks in the fitness tracker ecosystem and to summarize the valid consent criteria under the GDPR,
b) to review studies that have assessed the current privacy policies used by fitness app providers, in particular their consent management practices,
c) to evaluate heuristically the consent management of fitness providers that implement user consent by comparing their practices with the GDPR's criteria for valid consent, and
d) to summarize the issues identified in the reviewed studies that trigger the need for feasible solutions to fill the identified gaps.

### 1.2 Organization of this Review

This paper is organized as follows. In Section 2, we discuss the background knowledge and related works, present a literature review of the current privacy policies of fitness app providers, and explain the current legal frameworks and importance of compliance with these frameworks. In Section 3, we discuss the state of privacy used by fitness app providers and present our views based on the findings of assessments of the privacy of these wearable technologies. After a synthesis of the literature, in Section 4 we identify gaps in the literature and, in Section 5, we discuss feasible solutions that may help to fill the gaps identified. In Section 6, we conclude with a short discussion of the overall topic.

## 2 BACKGROUND

This section presents background knowledge and discusses the current state of knowledge about the privacy policies used by fitness app providers in the context of the data protection legal framework. It includes the authors' views and findings from several types of assessments on privacy policies used by these wearable technologies. We focus on understanding how data processing is bound to prior user consent and whether this consent is valid under the GDPR [1] criteria. Then, we describe our analysis of various issues that range from understanding the legal frameworks that protect fitness data to the state of privacy currently used by fitness apps. This review paper is organized around the following main topics.

**Legal frameworks for data protection and fitness data categorization** (Section 2). Fitness apps deal directly with individual users' information, such as health data, which are classified as highly sensitive data [10, 14, 31]. Hence, a legal framework is needed to protect users from unwanted disclosures.

**Current state of privacy policy used by fitness app providers** (Section 3). Researchers have raised serious privacy concerns about the use of data by fitness apps [2, 23–25, 30–32, 35, 47]. These concerns include linguistic assessment of the privacy policies of fitness apps, behavioral assessment of the privacy practice of fitness apps, and heuristic evaluation of consent practice related to fitness apps.

**Open issues, challenges, and recommendations** (Section 5). There is no uniform standard for the data format and content collected by wearable devices, and this lack of standardization may hinder methods to achieve consent for the management of these data. We discuss the existing research on consent management and system interoperability. Finally, we encourage future researchers to examine the feasibility of blockchain for addressing the problems identified.

In the following section, we explore the foundational concepts for the fitness tracker devices and legal frameworks for data protection. We then discuss the initial concept of fitness data categorization.

### 2.1 Legal Frameworks for Data Protection and Fitness Data Categorization

Different countries have various regulations and privacy requirements to protect an individual's data. Some of these regulations, including the GDPR [1], protect all forms of personal information, whereas others have limited scope, such as the Health Insurance Portability and Accountability Act 1996 (HIPAA) [33], which applies only to protected health information collected or held by an HIPAA-covered entity [50]. As more information moves to electronic form, increasing concern about privacy protection is expected [6]. Accordingly, several data-protection regulations and standards, including the HIPAA to Health Level 7 (HL7), have been introduced to address privacy concerns and to standardize electronic information [6]. However, these standards remain insufficient for meeting the various data types generated and rapid growth of wearable fitness devices [15].

Because they are considered "noncovered entities," fitness and health apps do not need to comply with the HIPAA, although they may need to achieve compliance once they interact with "covered entities" such as insurance companies and health-care providers [50]. The independence of designers, technologists, and policymakers has created a significant gap in which fitness app providers struggle to provide human-centered systems, achieve regulation compliance, and meet the technical aspects of their services all at once. Many experience noncompliance issues by designing systems with technological functions in mind while overlooking other aspects such as policy requirements relating to data protection.

Although law studies [9, 31] have argued and criticized the existing legal framework as ineffective for protecting individual health and fitness data from potential misuse, especially when gathered using wearable technologies, this paper focuses only on the technical aspects of the state of privacy within the existing legal framework.

This form of noncompliance is a new and challenging research avenue. Section 3 focuses on the associated issues that fitness tracker providers have with privacy preservation in processing individual data.

*2.1.1 Privacy Protection Through the GDPR.* The European Union's GDPR [1], which came into effect in May 2018, identifies six lawful bases for the processing of personal data to ensure that it is lawful, fair, and transparent. One of these legal bases, and the main focus of this review, is individual consent, which is considered valid under the GDPR [1] when it fulfills certain criteria; that is, consent must be unambiguous, informed, freely given, and specific [2]. Comprehensive summaries by Breen et al. [11] and Carvalho et al. [11] have identified all characteristics of valid consent under the GDPR [1]. For the purpose of consistency and clarity in this review paper, all related aspects of valid consent characteristics mentioned in the aforementioned studies [8, 11], and the relevant Consent number are summarized in **Table 1**.

*2.1.2 Privacy Protection Through the HIPAA.* Although other countries in the European Union have expanded protection to cover all individual data, recent law studies [29, 50] suggest that the regulations in the United States, such as the HIPAA, are inefficient at protecting fitness data. The HIPAA was passed by the US Congress in 1996 to expand health insurance coverage, reduce costs, and improve quality of health care in the United States. However, the HIPAA is considered to lack sufficient data protection because it applies only to "covered entities," which are defined as "health plans, health care clearinghouses and a health care provider who transmits any health information in electronic form" [29]. Accordingly, the US Department of Health and Human Services does not view fitness providers as covered entities unless the fitness provider works directly with a covered entity as defined under the HIPAA. For example, in the HIPAA, fitness data are not protected unless a person shares data with a physician, and then only that shared copy is protected by the HIPAA. Against this background, this review paper focuses on whether current device providers are compliant with the GDPR, which protects health *and* personal data, as defined in GDPR's Article 4 definition 1 as "personal data means any information relating to an identified or identifiable natural person" [1, p. 3].

## 2.2 Fitness Data Categorization

The consent criteria vary depending on the type of data being processed and whether it is sensitive. For example, in Article 9 in the GDPR [1], which discusses the processing of special categories of personal data, health data generated from wearable devices falls under "special categories." As noted by Mulder [31], this data category requires explicit $C7$ as well as standard informed consent $C2$ [1]. Colombo and Ferrari [14] have categorized the data type generated from wearable devices into four categories: identifier, quasi-identifier, sensitive, and generic. The authors derived these categories from privacy legislation, such as the GDPR [1] and data protection literature. These are depicted in **Figure 1** and **Table 2**.

Carminati et al. [10] applied Colombo and Ferrari's [14] four data categories to fitness internet of things(IoT) applications, as shown in **Table 2**. This categorization shows that the data for most fitness apps range from generic data to sensitive and personal data, which indicates the need for a strong data protection privacy policy.

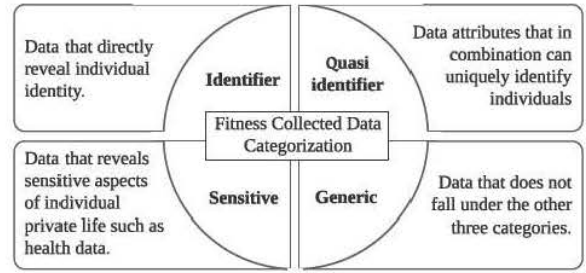

Figure 1: Data categorization [14, 48].

The consent criteria in the GDPR [1] aim to shift personal data control back to users, which creates a huge burden for the service provider [49]. For example, wearables such as Fitbit and Apple Watch rely on processing personal data, including sensitive information, as summarized in **Table 2**. The nature of these data indicate that implied consent, as in $C7$, is not sufficient but, instead, $C1$ consent must be obtained explicitly through "a clear affirmative action." Therefore, request implied consent is not sufficient and it must be explicitly given $C7$ through "a clear affirmative action" $C1$ [1, p. 10]. Fitness app providers must comply with regulations such as the GDPR [1] to achieve legitimate grounds for processing users' data and thereby avoiding future disputes. Even with the existing legal frameworks such as the GDPR [1] and HIPAA [50], when aiming to ensure transparency and privacy with the use of fitness devices, providers often neglect or misrepresent legal specifications, which has led to ongoing concerns about privacy and data [25]. The following section investigates the state of the privacy policies used by fitness app providers.

## 3 STATE OF PRIVACY POLICIES USED BY FITNESS APP PROVIDERS

This section discusses the state of the privacy policies used by providers of fitness apps and the importance of transparency, which is often considered a core element for achieving both privacy and compliance. The GDPR [1] and the Australian Privacy Act 1988 [34] foster transparent data processing practices to enhance trust in privacy protection. In technical terms, Seneviratne and Kagal [43] developed privacy-enabling transparent systems that rely on transparency as a critical component, in which all processes are audited. To ensure that there are no policy violations, the subject is given a clear view of how their data are used. Seneviratne and Kagal's [43] approach and the legal frameworks stress the importance of transparency in any system. Hence, we define the transparency of the system based on three fundamental pillars, as shown in **Figure 2**.

Thus, for fitness app providers to establish legitimate processing grounds in this context, a consent management mechanism that complies with the GDPR [1] is required to improve the transparency of the data flow and to control whether consent has been granted, denied, or revoked. To ensure data flows in a privacy-preserving manner, user consent can be integrated into the processing [6, 38]. Therefore, a solution is needed that ensures coverage of all fundamental pillars of transparent system by complying with the GDPR and integrating consent to preserve privacy. The next section focuses on the issues associated with privacy policies used by fitness apps.

| Consent number | Consent characteristics | Definition | GDPR's Recital | GDPR's Article |
|---|---|---|---|---|
| C1 | Unambiguous | Preselected options or opt-out requests are invalid, as consent must be through "a clear affirmative action"; i.e., the practices of ticking the box to agree to terms and conditions is problematic. | 32 | - |
| C2 | Informed | The data subject must be aware of all information associated with processing their data before the data processed. This also requires this information should be understandable using a clear and plain language. | 32, 42 | 7, 2 |
| C3 | Freely given | Consent should be given on a voluntary basis. Data subject must be aware of all consent effects. Consent presumed to be invalid if there is any coercion or obligation. i.e., require non-negotiable consent. | 32, 42, 43 | 7, 4 |
| C4 | Specific | The consent request for processing data must be granular and the purpose must be specified. Thus, data subject must be fully aware of the intended purposes and the used method to process their data. Hence, an opt-in option should be given for each purpose separately instead of a blanket consent. | 32 | 7, 2 |
| C5 | Auditable | All consent requests, grants, and revokes; and other details, must be stored for future auditing which can be used later as valid proof. | 42 | - |
| C6 | Withdrawal | Consent request should also include the method to withdraw from granted consent and it should be easy to withdraw. | - | 7, 3 |
| C7 | Explicit | Consent should verify that consent was given by the data subject, and it should include detailed information on what data consented, thus explicit consent must be verifiable to be validated. | 51, 71 | 9, 2 |

**Table 1: A summary table of valid consent characteristics under the GDPR [1, 8, 11].**

| Data category | Definition | Fitness data |
|---|---|---|
| Identifier | Data that directly reveal individual identity. | • Name<br>• Location<br>• Phone number |
| Quasi identifier | Data attributes that in combination can uniquely identify individuals. | • Birth date |
| Sensitive | Data that reveals sensitive aspects of individual private life such as health data. | • **Health:** Physical state, blood pressure, heart rate, weight, height, and psychological state.<br>• Position |
| Generic | Data that does not fall under the other three categories. | • **Fitness:** Walking and running steps |

**Table 2: A summary of fitness data categorization [10, 14, 48].**

### 3.1 Linguistic Assessment of Privacy Policies Used by Fitness Apps

Research into fitness apps have revealed a gap between users' understanding of consent and subsequent data usage by service providers [25, 31, 47]. Thus, consent given to privacy policies must be written in clear and plain language to achieve transparency.

Although transparency is a crucial when processing personal data, the assessment of Sunyaev et al. [47] of the privacy policies of the top 300 health apps in the Apple and Android app stores provides contrary evidence. They found that the sharing practice of individual personal data on these health apps is far from transparent and that, because of the lack of clarity, the overwhelming amount of text makes it difficult for users to find and understand the privacy policies.

A recent law study by Mulder [31] examined the privacy policies of 31 health apps, including fitness tracking apps, such as Fitbit, Strava, and Nike Running, and compared these with the GDPR. The authors focused on identifying noncompliance in the policies and the legal consequences. The findings suggest that the key reasons for users' lack of knowledge about the consent process is that policy statements are lengthy and use complex language [31]. Failure to state clearly the purpose of processing personal data violates the GDPR criteria, which state that consent must be informed $C2$ and, if not, it is unlawful to process the user's data [1]. These findings suggest the need to adopt a more transparent system that clearly communicates the purpose of data requests.

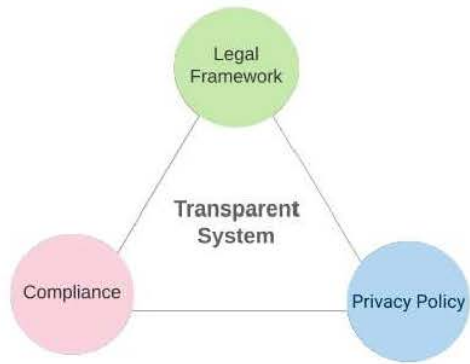

Figure 2: Fundamental pillars of a transparent system.

### 3.2 Assessment of the Behavior of Privacy Practices Used by Fitness Apps

In addition to linguistic assessments, other types of assessments have focused on privacy risk analysis of the behavior of health and fitness apps [2, 23, 24, 30, 32]. A descriptive study by Grundy et al. [23] used network analysis of data-sharing links among fitness apps and found significant privacy concerns because of the central position of fitness apps in mHealth data-sharing ecosystems such as Jawbone UP, Apple HealthKit, Fitbit, and Strava. The high interconnectivity of these apps indicates the potential for the sharing of user data with third parties in an unanticipated and involuntary manner [23]. However, this analysis did not distinguish whether fitness apps collected user data passively or through prior agreement with users.

Further analyses have been conducted to assess the privacy of sharing user data based on traffic analysis and user review [24, 30]. For example, using the proposed system "AntMonitor" of Shuba et al. [24], Hatamian et al. [24] performed traffic analysis to detect any leaks of sensitive data to a remote server without the user's consent. This traffic analysis examined whether there was data transmission to a third party when the user was not interacting with the device [24, 45]. Using their standalone "MARS analysis," these authors also analyzed the correlation between the user's perception of privacy and the real behavior of health and fitness apps. Hatamian et al. [24] confirmed that current privacy practice among fitness apps is problematic in terms of accessing sensitive data without transparent reasons.

Momen at al. [30] extended the work of Hatamian et al. [24] by comparing privacy-related complaints against fitness apps from before to after the GDPR went into effect. Most of the complaints related to overprivileged apps that request permission unrelated to their functionality. The findings of Momen et al. [30] suggest a positive impact of the GDPR on app behavior. Based on the ethical principles of autonomy and competence, Ahmed et al. [2] used the conceptual model of informed consent proposed by Friedman et al. [19] to analyze the status of consent management in online social networks. Ahmed et al. [2] found that a common practice among service providers is based on Hobson's choice; that is, data subjects are required to agree entirely with all privacy policies or be prohibited from using the service. Similarly, Neisse et al. [32] found that most apps use end-user license agreements as a convenient way to seek user consent, which is not always practical in the current use of wearable devices. This form of one-time consent is not truly voluntary or generic and, therefore, it violates the criteria that consent is freely given $C3$ and specific $C4$; in addition, users should not be affected if they decide not to grant permission [1].

These behavior assessment findings suggest that fitness apps share user data with other entities involuntarily. This is because of the unclear one-time consent used to obtain user consent and the lack of the user's right of withdrawal. These issues should be addressed by a new system that records all exchanged data activity while requiring with user consent.

### 3.3 Heuristic Evaluation of Fitness App Consent Practices

The previous noncompliant practices were for fitness apps that use one-time consent to obtain users' consent to process their data [2, 32]. Some fitness apps have moved from traditional one-time consent to more flexible management of user consent. This subsection critically compares the privacy policies of the three top fitness apps shown in **Table** 3 (Fitbit, Apple Watch, and Strava) to understand their consent culture with more specific reference to the criteria for consent validity under the GDPR (Table 1). The qualitative approach used in this evaluation is based on the heuristics framework of Hutton et al. [25] (Figure 3). We examined the state of privacy in these fitness tracking apps using this framework, which is used to assist both researchers and data controllers to conduct impact assessments for both privacy and data protection.

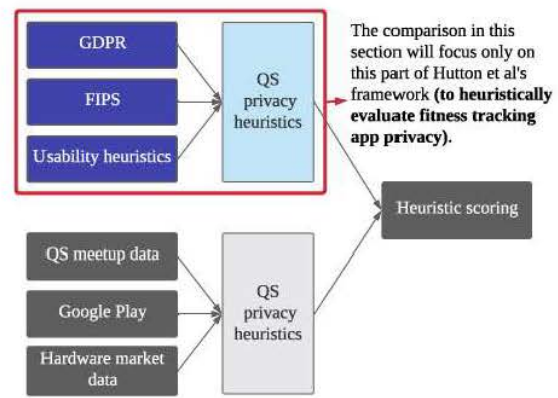

Figure 3: Heuristic framework of Hutton et al. [25].

Our pilot evaluation of the fitness app consent practices included in **Table 3** showed that none of the three fitness app providers offer a fully dynamic consent mechanism. Although they give the user the right to withdraw, they do not update this consent when new data are collected or if purpose of the request changes. Moreover, these apps do not use one-time consent and do not give the user the right to choose which data can be processed, which is important because wearable devices generate different types of data, some of which are sensitive. For the providers of fitness apps, avoiding noncompliance is a rational decision and is more cost-efficient than dealing with the consequences of unwanted disclosures of personal data [47]. Devising a consent model is an important step in ensuring compliance with the GDPR and helping fitness app providers to meet privacy requirements related to the processing of individual data [35]. Therefore, a solution that uses a dynamic consent mechanism to address these issues is required.

## 4 SUMMARY OF THE STATE OF THE ART

This section summarizes the current understanding of the privacy policies of fitness apps and the gaps identified in this review. This review paper started by discussing the legal frameworks for privacy protection and summarizing the valid consent criteria under the GDPR, and then described the importance of adopting legal frameworks in the fitness-tracker ecosystem. This summary of valid consent criteria used throughout this review paper was used to investigate compliance issues related to consent management practice. We have reviewed the current state of privacy policies used by fitness app providers, particularly their consent-management practices. This review has discussed three types of assessment of fitness app privacy policies: linguistic assessment, behavior assessment, and heuristics evaluation.

We have identified the research gaps in the area, such as lack of system transparency, lack of privacy policy legibility, privacy concerns associated with one-time consent, and noncompliance with the adopted consent management solution. Thus, our identified problems and its relevant number, are stated as follows:

**P1** *Lack of system transparency.* Fitness app privacy policies lack transparency in their consent practice. To overcome this issue, consent management mechanisms that comply with GDPR [1] are required. To ensure that data flow in a privacy-preserving manner, user consent should be integrated [6, 38]. A solution that ensures that all fundamental pillars of a transparent system that comply with the GDPR and integrate consent to preserve privacy is needed.

**P2** *Lack of fitness app privacy policy legibility.* The findings from linguistic assessment of fitness app privacy policies suggest the need to adopt a more transparent system that clearly communicates the purpose of requesting data. A solution that enhances the consent's legibility by ensuring consent is obtained using plain language that includes the attributes and detailed information related to the purpose of each data request separately is required.

**P3** *Concerns with one-time consent.* The findings from behavior assessments suggest that fitness apps share user data with other entities involuntarily. This occurs because of the unclear one-time consent required of users and the lack of their right to withdraw consent. Therefore, it is crucial to address these issues by proposing a system that records all exchanged data activity tied to user consent. A solution that allows the user to withdraw or update consent at any time is needed.

**P4** *Noncompliance issues with new consent management.* Although some fitness apps have moved from traditional one-time consent to flexible consent, noncompliance issues remain. Our findings from the heuristics evaluation suggests fitness apps do not offer a fully dynamic consent mechanism, nor do they update consent when new data are collected or if the purpose of the request changes. They also do not give the user the right to determine which data can be processed. A solution that identifies specific data to be shared, as well as not shared without user consent, is required.

In Section 5, we review the current solutions to address the aforementioned gaps in different domains in terms of consent management; these include open issues, challenges, and our recommendations.

## 5 OPEN ISSUES, CHALLENGES, AND RECOMMENDATIONS

In this section, we discuss the potential of existing solutions to address the problems identified as **P1**, **P2**, **P3** and **P4** in Section 4, including but not limited to centralized and decentralized authority, and distributed technology. Given that consent is defined as the method for asking for the user's permission to read and edit their health records [6], the solutions must address the privacy issue by obtaining the individual user's consent.

### 5.1 Open Issues and Challenges:

*5.1.1 Centralized Authority.* Bacchus [6] has presented a novel protocol for managing consent requesters, known as consent-based access control (CBAC). This protocol aims to switch the control of patients' sensitive health data back to their original owner, the patient. This suggested CBAC approach involves, first, negotiating with the patient to gain access to their health data and then, once an agreement has been reached, the data requester is issued a consent token that allows access to that information. Although this proposed approach helps to solve the lack of patient control over access to their health records, it remains a one-time consent management **P3** and it places the ultimate trust in a single authority **P1**. There are risks associated with trusting a single authority, such as malicious acts, lack of transparency, and forged consent. This approach also means that users must blindly trust a single authority to handle their data rather than trusting the technology behind it. By contrast, implementing a system that uses distributed technology rather than a centralized authority can resolve trust issues. Blockchain technology introduces many built-in features that make this technology a suitable candidate for consent management. Subsection 5.1.2 discusses the potential of this distributed technology to solve issues **P1**, **P2**, **P3** and **P4**. Thus, now turn to the discussion of blockchain solutions for consent management and interoperability issues associated with fitness apps.

| Category description | Heuristic | GDPR | Fitbit | Apple watch | Strava |
|---|---|---|---|---|---|
| **Choice or Consent** | | | | | |
| H8 | Consent acquired before data shared with remote actor. | C2 | × | × | ✓ |
| H9 | Consent is explicitly opt-in: no pre-ticked checkboxes, etc. | C1 | ✓ | ✓ | ✓ |
| H10 | Can choose which data types are automatically collected from sensors or other sources, for example, connect a finance app to a single bank account or track steps but not heart rate. | C4 | × | × | × |
| H11 | Data collection consent is dynamic: if new types of data are being collected, consent is renewed in situ. | C6, C2, C4 | × | × | × |
| H12 | Data processing consent is dynamic: if the purpose of processing changes, consent is renewed. | C6, C2, C4 | × | × | × |
| H13 | Data distribution consent is dynamic: if the actors' data are distributed changes, consent is renewed. | C6, C2 | ✓ | × | × |
| H14 | Consent to store and process data can be revoked at any time: with the service and any other actors. | C6 | ✓ | ✓ | ✓ |
| H15 | Can control where data are stored. | - | × | × | × |
| **Additional consent characteristics check** | | | | | |
| H16 | keep record of all consent processes (demonstrating consent) | C5 | × | × | × |

Table 3: Comparison of GDPR and the privacy policies of Fitbit [18], Apple Watch [3] and Strava [46] in the choice and consent category [25] and auditing criteria.

*5.1.2 Blockchain and Interoperability Issues.* Although the studies described in Section 3 highlight the importance of adopting a transparent consent mechanism in any system to address noncompliance issues, many have discussed the lack of interoperability between wearables platforms [2, 23–25, 30–32, 35, 47], which creates a huge barrier to their widespread use and limits the exchange of personal data with other entities [15, 26]. The problem relates to the great heterogeneity of fitness data representations that remain confined to fitness provider platforms, which makes the integration of these data silos with other entities difficult [26, 37].

Research studies have recommended joint action across the wearable ecosystem to overcome this interoperability issue [17, 20, 37, 42]. Saripalle [42] proposed a solution for integrating fitness data with other entities such as electronic health records (EHRs) by leveraging the HL7 Fast Healthcare Interoperability Resources (FHIR) standard [42].

We note that the FHIR standard is currently used in some research blockchain designs in the health-care context. Rupasinghe et al. [38] used an FHIR consent resource in their proposed design to avoid integration failures with EHRs. According to their literature review, only five of 23 proposed solutions addressed interoperability issues in their consent management [5, 16, 27, 51]. Additionally, some health apps such as Apple's Health Records app use FHIR to allow users to download data from their health-care providers [4]. Although designing a system in an interoperable manner is important for being able to combine fitness data, there is still no common standard for consumer fitness devices [15]. This is because interoperability standards go through a series of approvals before being piloted, but wearable fitness devices go to market faster than the relevant standards are developed [15]. The capacity to share individual personal data depends on obtaining the individual's consent, and this in itself is hindered by interoperability issues.

## 5.2 Recommendations and Future Work

To improve users' control over the processing of their fitness data while simultaneously enabling fitness tracker providers to comply with the GDPR obligations, we have identified and investigated past and existing privacy-preserving practices of fitness apps to suggest suitable solutions for the problems identified. As an emerging technology, blockchain has outstanding advantages for resolving these consent-management privacy concerns. We propose a blockchain-based consent mechanism, which we suggest can improve privacy by designing a human-centric and legally compliant system that manages user consent dynamically around the sharing, collecting, and processing of fitness data between the requester and user under the GDPR's validity criteria. By leveraging the problems identified as **P1**, **P2**, **P3** and **P4** (Section 4) from our review of the literature and our discussed solutions (Section 5.1), we encourage future research to investigate empirically our proposed fitness app consent-management requirements (as listed below) through the lens of the advantages of blockchain:

- **R1** *Trust or security*
- **R2** *Transparency*
- **R3** *Scalability*
- **R4** *Auditability*
- **R5** *Preservation of the original functionality*
- **R6** *GDPR compliance*

Based on our findings (Section 5.1) of the existing solutions for preserving user privacy and blockchain's potential to address the problems identified as **P1**, **P2**, **P3** and **P4** (Section 4), we recommend further research to examine the feasibility of using blockchain to manage user consent through some or all of our recommended requirements **R1**, **R2**, **R3**, **R4**, **R5** and **R6**. Researchers might manage user consent by leveraging smart contracts to automate the processes of consent grants and revoke actions. In addition, blockchain and smart contracts might also be used as tools to detect misbehavior by fitness providers if they process or share data in an unanticipated and involuntary manner.

## 6 CONCLUSION

This review paper aimed to identify the importance of adopting legal frameworks in the fitness-tracker ecosystem by reviewing data protection legal frameworks and the relevant literature, and by identifying the problems inherent in the compliance with existing legal frameworks of the privacy policies of fitness apps. The studies reviewed included linguistic, behavioral, and heuristic assessments of the consent practices of fitness apps. From our analysis, it is clear that there are serious privacy concerns associated with the use of fitness apps. Our analysis identified four problems related to the preservation of privacy of fitness data: lack of system transparency, lack of privacy policy legibility, concerns regarding one-time consent, and issues of noncompliance regarding consent management. Identifying these problems enabled us to review and discuss the existing solutions including the centralized and decentralized authority that might be feasible for addressing these identified gaps and improving the user's control over the processing of their fitness data. Blockchain's built-in features make it a suitable candidate for solving the problems identified. However, further research is needed to examine the feasibility of using blockchain to manage user consent.